# Supply Chain Management analysis: a simulation approach of the Value Chain Operations Reference model (VCOR)

**Yacine Ouzrout** *
LIESP-Lyon2 Laboratory, Université of Lyon 2,
160 Bd de l'université 69676 Bron Cedex, France,
E-mail: yacine.ouzrout@univ-lyon2.fr
Phone: +33.478.774.484
**Corresponding author*

**Matteo M. Savino**
Department of Engineering, University of Sannio,
Piazza Roma, 21, 82100, Benevento, Italy,
E-mail: matteo.savino@unisannio.it

**Abdelaziz Bouras**
LIESP-Lyon2 Laboratory, Université of Lyon 2,
160 Bd de l'université 69676 Bron Cedex, France
E-mail: abdelaziz.bouras@univ-lyon2.fr

**Carlo Di Domenico**
Hermes Reply S.p.a,
Corso Francia, 110, Torino, Italy.
E-mail: didomenico@hermesreply.com

## 1. Introduction

Scientific literature related on Supply Chain Management (SCM) is rich of publications, but the reality is that there is a lag between practice and theory (Balan, et. al, 2006; Simchi-Levi et al, 2000; Neubert et al, 2004). Mass media and Internet have speeded up the diffusion of new products; at the same time, technical innovation and market severe competition promote rapid obsolescence of existing products and technologies. When a company succeeds in developing a new product category, other competitors may soon emerge. The market originator must endure not only the substantial risk of whether the market would materialize or not, but also the difficulty of recovering major costs, such as research and development and advertisements. Increasingly, the supply chain becomes the mechanism for coping with these problems because it is often inefficient for any single company to produce a whole product (Abdullah et al., 2004; Feller et al, 2006).

Hence, modern business is essentially the competition of one supply chain (SC) with another. SC dynamics is the interaction processes of the participants from different departments and companies. A positive aspect of supply chain dynamics is effective collaboration, which may lead to higher performance. A negative aspect is independent decision making, which may create various delays and aggravate forecasting errors (Simatupang and Sridhran, 2002; Abu-Suleiman et al. 2005).

This research stresses the interest on the possible quantitative advantages given by the introduction of the *Value Chain* concept into the SCM through a simulation approach. Discrete event simulation, continuous time-differential equations, discrete time difference models and operational research techniques are some of the commonly used quantitative modelling techniques to evaluate and design supply chains (Lee et al., 2002; Terzi and Cavalieri, 2004). The correct identification of key variables and their interactions, together with determining how the information can be better managed enables the utility of the entire supply chain.

In this work we use discrete event formalism to model and study the supply chains. Discrete-event simulation is chosen for its capability to represent physical and information flows along with their respective delays, in an information feedback control type of setting. Our main research interest is to clarify the critical factors for minimizing the negative effects of supply chain dynamics and to gain insight on how to effectively manage them. To achieve these objectives we developed a simulation model to implement the VCOR and to verify the possible advantages that it is able to obtain.

## 2. The Value Chain management theory

A Supply Chain can be defined as a system network that provides raw materials, transforms them into intermediate commodities and/or in finished goods and distributes them to the customers through a delivery system (Christopher et al, 2002). The aim is to produce and distribute the right quantities, to the right locations, at the right time, while reducing costs and maintaining a high level of service. SCM is concerned with smoothness, economically driven operations and maximising value for the end customer through quality delivery. The limitations are however due to the fact that SCM as a concept does not extend far enough to capture customer's future needs and how these get addressed, and furthermore, it does not encompass the post-delivery, post-evaluation and relationship building aspects (Al-Mudimigh et al, 2004). Another important theory can be defined as strategic in the context of SCM, the concept of value chain management.

The Value Chain was described and popularized by Michael Porter in his best-seller, Competitive Advantage: *Creating and Sustaining Superior Performance* (Porter, 1985). Porter defined *Value* as the amount that buyers are willing to pay for what a firm provides, and he conceived the *Value Chain* as the combination of generic activities operating within a firm, activities that work together to provide value to customers. Porter linked up the value chains between firms to form what he called a *Value System*. However, in the present era of greater outsourcing and collaboration, the linkage between multiple firms' value creating processes has more commonly become called as Value Chain. As this name implies, the primary focus in value chains is on the benefits that accrue to customers, the interdependent processes that generate value, and the resulting demand and funds flows that are created.

Feller, Shunk, and Callarman (Feller at al. 2006; Jan Olhager et al, 2006) exposed some important considerations about the value concept. First is that value is a subjective experience that is dependent on context. The same product or service has not the same value in different parts of the world or in different situations. Second, value occurs when needs are met through the provision of products, resources, or services. Finally, value is an experience and it flows from the customer. Clemmer (1990) affirms that customer value is layered and has been described by three concentric rings. In the center ring is the product value which is the technical value derived from providing a source of supply. A second ring of service value is provided by the services that surround the product such as personal care and warranty service. The third ring has been called the new service/quality battleground, and was made popular by business thinkers such as Peters and Waterman ("In Search of Excellence"). This third level of value is achieved by providing enhanced service, to "make your customer successful" rather than just satisfied.

The huge importance of focusing on the customer has forced the integration of the optimization techniques of the Supply Chain Management, Customer Relationship Management (CRM) and Product Lifecycle Management (PLM), as shown in figure 1.

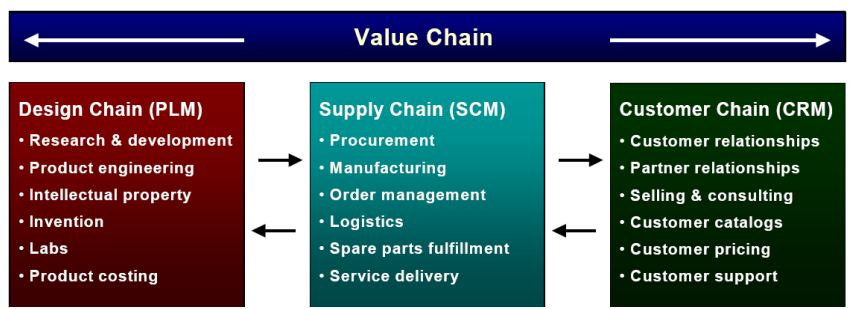

Figure 1 - Integration of PLM, SCM, and CRM from www.process-wizard.com

The value chain theory focuses on the management of the product lifecycle (PLM) which is a strategic business approach that helps enterprises to achieve its business goals of reducing costs, improving quality and shortening time to market, contemporarily innovating its products, services, and business operation (CIMdata, 2002). By increasing an enterprise's flexibility to respond quickly to changing market pressures and competitors, the value chain helps companies to:
- deliver more innovative products and services;
- Reduce costs;
- Improve quality;
- Shorten time to market while achieving the targeted Return On Investment (ROI);
- Establish more collaborative, and improved relationships with their customers and suppliers (Garetti et al, 2005).

The value chain management includes also the Customer Relationship Management (CRM) which consists on the creation, the development, the palimony and the optimization of long period relationships more profitable among consumers and firm. The success of CRM is based on the understanding of the consumers' needs and desires, and it fulfils setting such desires to the center of the business, integrating them with the firm's strategy, the people, the technology and the business process (Munari, 2004).

## 3. Supply Chain and Value Chain Modeling

The first general framework for SCM, the Supply Chain Reference Model (SCOR) was developed by the Supply Chain Council (SCOR, 2005); the model is rather general, defining the supply chain standard processes and establishing standard terminology in quite broad terms. SCOR spans customer and market interactions and the physical material transactions. This model can also help manufacturers to carry-out benchmarking against other well-established companies, for that the model proposes some best practices and Key Performance Indicators (KPI). The SCOR-model has been developed to describe the business activities associated with all phases of satisfying a customer's demand; it consists of a plan, source, make and deliver process elements (level 1 of the model) which revolve around the entire supply chain. The main assumption of the model is that by integrating the process elements along the supply chain, companies should become more competitive. But the support functions such as administration, R & D, and customer services are not included (Bolstorff and Rosenbaum, 2003).

Nowadays manufacturing in Europe is going on a deep change. The increase of the productivity has concentrated the attention on the approach for the achievement of competitive advantage through efficiency actions and process optimization. Besides, the European enterprises have realized that a complementary kind of initiative is necessary for the product design and engineering processes. It is indispensable, in fact, to keep a better know-how in order to remain competitive and to be able to offer new and advanced products to the market. The introduction of concepts like PLM has become essential to acquire new clients and new market segments, and to adopt concepts of Value Chain Management all over the network.

In late 2003 and early 2004 a series of meetings culminated in the development of the first iteration of the Value Chain Operations Reference (VCOR) model. Participants in these meetings came from a global pool of business process knowledge experts many of whom worked for large end-user, consulting or software companies, domain specific not-for-profit organizations, or academia; they created the Value Chain Group: VCG (Value Chain Group, 2005). The VCOR model is able to achieve some benefits to firms summarized as follows:

- A standard based approach to define essential collaborations between trading partners.
- Agreement on product life cycle objectives and how to achieve them.
- Reusable process templates based on best practices.
- Integration of existing and new information management systems.

- A fast response to changes while maintaining and extending value chain performance.

### 3.1. The VCOR Framework

The VCOR model was developed from the perspective of being a value chain framework with the development of Seven Performance Attributes linking the three domains of Product Development, SCM and Customer Chain across the supply networks together. The structure of VCOR model supports and enables companies to integrate their three critical domains: Product Developments, Supply Network Integration and Customer Success, using one reference model to support the vision of an integrated value chain (Heinzel, 2005). To achieve this goal VCOR uses a "process based, common language" of syntax and semantics while, at the same time, create a base for the successful Service Oriented Architecture Game Plan. The main objective of this model is to increase the performance of the total chain and support the current evolution; for that, it proposes four different modeling layers:

Strategic Level: The Top Level of the model includes all the high level processes in Value Chains and is represented through the Process Categories Plan – Govern – Execute. The Level is defined to be the Strategic Level of the Model, meaning that this is where high level decisions are made regarding how to gain a competitive advantage for the Value Chain in scope. The VCOR Strategic Level has three Macro-processes: *Plan:* it balances the current strategic objectives with current asset status and produces decisions on activities to move the organization toward the goals; *Govern:* decision based process which identifies and enables a value chain by establishing the rules, policies and procedures to control the implementation of Plan and Execute processes within the Value Chain; *Execut*e: it transforms the customer requirements to production processes. The Execute Processes operate within the limits of the Management criteria and to the parameters defined by the Plan Processes.

Tactical Level: The second level of the model contains "abstract" processes decomposed from the Strategic Level to implement and fulfill the strategic goals set in the top level of the model hierarchy a set of tactics needs to be developed and realized. The Tactical Level can be described being instituted for "Horizontal Value Chain Process Re-Engineering". The VCOR model processes decomposes from Strategic to the Tactical Level with Plan and Govern, keeping their respective naming in the first part of the process notations on this level as they influence each of the Execution Processes.

The figure 2 illustrates the decomposition of the strategic level into component processes. The Market Process has deep impact on the entire chain and for this reason is extended over the diagram. The Plan and Govern processes have the same name (e.g. Plan Research, Govern Develop and so on). To describe these different processes, VCOR defines three main groups: Market, Research & Develop, Acquire, Build & Fulfill, and Sell & Support.

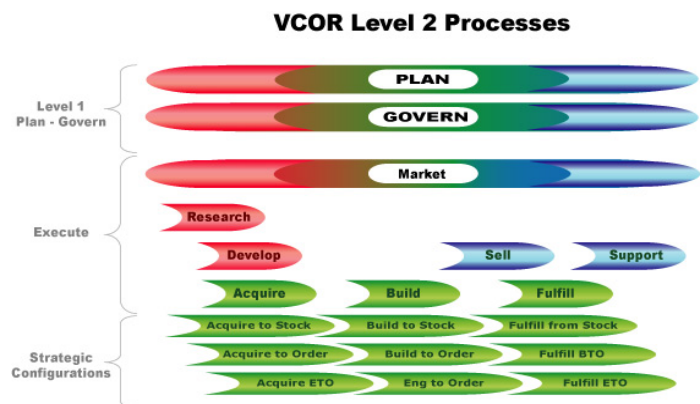

Figure 2- VCOR Layers Diagram from EMEA-VCG

Operational Level: The third level of the model represents specific processes in the value chain related to actual activities being executed clarifies. On this level focus is usually vertical business process improvements or business process re-engineering as usually it is named. In a value chain perspective this is the level where fine-tuning occurs.

### 3.2. VCOR vs. SCOR: Metrics and Performances

In order to measure the performance of the chain, the Supply Chain Council has introduced five metrics (see table 1) and KPI (Key Performance Indicators) in the first level of the SCOR model to test supply chain reliability, responsiveness, flexibility, costs and efficiency in managing assets (working and fixed capital). For the remaining processes (level two) and elements (level three) there are specific KPI to test the performance of each part of the SC configuration. Moreover, for each process, some best practices are defined to simplify the analysis of the chain. These are reliable points of reference, given by the experience of enterprises leader in their own field, to follow in order to improve the performance without trying unpredictable strategies that could be dangerous. A company cannot be best in all metrics, so it should wisely target its strength and differentiates itself in the market, while ensuring that it stays competitive in the others (Bolstorff and Rosenbaum, 2003; Hausman, 2002) Actually, the most part of companies do not choose to improve all indicators but they focus on some of these, building their strength points. According to the definition of Value Chain Group, a metric is *"a quantifiable variable that reflects a specific state of business performance during process execution within a strategic value chain context"*. In VCOR a metric is characterized by different features: Metric Name, Metric Definition, Priority Dimension, Metric Class & Sub-Class, Formula, Input Requirements, Dimension, Calculation Rules etc. The table 1 describes the definition of the seven dimensions defined in the VCOR model and compares them to the SCOR's model definition.

**Table 1** SCOR strategic performance metrics vs. VCOR priority dimension

| | | |
|---|---|---|
| **Delivery/ Reliability** | *SCOR* | *Definition:* The performance of the SC in delivering: the correct product, to the correct place, at the correct time, in the correct condition, packaging and quantity, with the correct documentation, to the correct customer. *Metrics:* Delivery performance; Fill rates; Perfect order fulfillment. |
| | *VCOR* | *Description:* The ability to deliver the correct product to the correct market and customer on time. *Metrics:* Deliver Performance; Request Date; Product Release Variance; Forecast Accuracy. |
| **Responsiveness/ Velocity** | *SCOR* | *Definition:* The velocity at which a SC provides products to customer. *Metrics:* Order fulfillment lead times. |
| | *VCOR* | *Description:* The cycle time taken to deliver a product or service to the customer. *Metrics:* Order Fulfillment Lead Time; Product Development Lead Time. |
| **Flexibility/ Adaptability** | *SCOR* | *Definition:* The agility of a supply chain in responding to marketplace changes to gain or maintain competitive advantage. *Metrics:* SC response time; Production flexibility. |
| | *VCOR* | *Description:* The capability in responding to market change to gain or maintain competitive advantage. *Metrics:* Delivery Adaptability; Value Ch. Agility; Ideation Yield. |
| **Costs** | *SCOR* | *Definition:* The costs associated with operating the SC. *Metrics:* Cost of goods sold; Total SCM cost; Value-added employee productivity Warranty/return processing costs. |
| | *VCOR* | *Description:* The cost associated with operating a value chain. *Metrics:* Cost of Quality, Design Cost Ratio; Logistic Cost Ratio; Manufacturing, Sales & Marketing Cost Ratio. |
| **Effectiveness/ Asset** | *SCOR* | *Definition:* The effectiveness of an organization in managing assets to support demand satisfaction (Fixed and working capital). *Metrics:* Cash-to-cash cycle time; Inventory supply; Asset turns. |
| | *VCOR* | *Description:* The effectiveness of an organization in managing assets of the Value Ch. to support market and customer satisfaction. *Metrics:* Asset Turnover; Cash Conversion Cycle; Design Realization; Inventory Supply. |
| **Innovation** | *SCOR* | *Definition:* -  *Metrics:* - |
| | *VCOR* | *Description:* The ability to strategically leverage internal & external sources of ideas and introduce them to market through multiple paths. *Metrics:* Product Innovation Index; R&D Profit Contribution. |
| **Customer** | *SCOR* | *Definition:* -  *Metrics:* - |
| | *VCOR* | *Description:* The capability to develop positive collaborative customer relationships. *Metrics:* Customer Growth Rate &Retention Rate, Market Share. |

*3.3. VCOR vs. SCOR: Synthesis*

The two models have been created in order to give a reference point to those companies that applied the principles of SCM and Value Chain Management. The SCOR is an affirmed and reliable model and has preceded the VCOR by few years. For this reason they have some analogies inherent in the structure and differences in their scope. Both models tried to be standards, improved by Supply Chain Council and Value Chain Group respectively, to simplify their correct realization in real cases and facilitate the communication between trading partners. Moreover, they are based on high-level generic process categories to fit better for all types of firms. But the feature that perhaps is the reason of their success is the capability to calculate the level of performance for every single process of the chain in a very simple way and, at the same time, to introduce best practices to facilitate the management of productive processes. It is crucial to found strategic choices on objective data that can be compared with the metrics filed during years of activity or with performance indicators of other companies if possible. Differences between the SCOR and VCOR are that VCOR is an enterprise model using a framework and taxonomy that features governance and the decision making processes at its highest levels with interconnectivity to all Enterprise processes. The interaction of process elements in the supply chain domain are mostly "transactional" as opposed to interaction of Enterprise business process elements involving higher order information processing in the decision making process.

A question has arisen: does a company that successfully implements the SCOR model really needs to change framework facing other efforts? The answer is not trivial, but it is possible to state that the paybacks gained with the VCOR can benefit an enterprise enough to precede its own competitors, extending its market. Furthermore, the extension from SCOR to VCOR is not very intricate because VCOR's basic design uses the framework and methodology of a unified General Systems model that can be applied to most organizational type (Feller et al, 2006). Expansion occurs by reaching out into existing vertical domains of SCM, PLM and CRM to integrate their respective business process elements into a unified Value Chain model similar to SCOR. In this sense the VCOR can be considered the natural extension of the SCOR model.

In the next sections we will describe a simulation model realized to allow companies to evaluate in advance the possible advantages of the Value Chain implementation. This model includes some of the metrics and KPI presented in the first part of the paper.

## 4. The simulation of the Value Chain processes

In this section we front the problem of the development of a simulation architecture for the implementation of a VCOR model. Nowadays in literature there are few examples of global simulated Supply Chain because it is no easy to model the entire chain and implement it by specific and dedicated software (Lee et al, 2002; Kleijnen, 2005); moreover the number sensitively increases if we consider the examples that apply the SCOR (Boucher et al, 2003, Jain and Leong 2005, Herrmann et al, 2003).

This section presents first the advantages of the simulation in comparison to the mathematical models and finally the realization of a generic architecture to simulate the VCOR model. Many researchers tried to solve the production-distribution issues in the form of analytical problems. The objective in this type of method is to minimize the overall production and distribution costs in multi-facility, multi-product and multi-period problems. The algorithms are mainly based on heuristics or on network flow; but it is obvious that the complexity of the problem grows in function of the number of elements in the chain: number of products, resources, constraints... if we also introduce the uncertainty the modelling and the problem solving became prohibitive.

In this context of dynamic, stochastic and complex systems it is difficult to analytically model the problem, but the limits of the mathematical analysing can be solved by computer simulation. The benefits in using simulation in Supply Chains can be summarized as follow (Kelton et al, 2004):

- Capacity to capture data for analysis: users may model unexpected events in certain areas and understand the impact of these to the SC

- It can decrease drastically the risks inherent to changes in planning: users may test several alternatives before making the planning change.

- Investigating the impact of: innovations within the SC; eliminating an existing infrastructure or adding a new one within the SC; strategic operational changes, such as process, location and use of new facilities, the fusion or the separations of some components of two SC; manufacturing products inside the company; creating new suppliers or subcontracting some processes;

- Investigating relations between suppliers and other components of the Supply Chains to rationalize the number and size of order lots, using as a basis the total of costs, quality, flexibility and responsibilities;

- Investigating the opportunities to decrease the varieties of product components and standardize them throughout the Supply Chains.

These main reasons have lead us to choose a simulation approach to analyse the SC dynamic and behaviour and to define a simulation model to implement the VCOR processes. The next paragraphs describe the generic supply chain simulation model that we propose; from the configuration to the description of the different blocks in ARENA.

### *4.1. Model Implementation*

The objectives of this section are to describe the VCOR model simulation architecture and to illustrate the implementation of its different elements.

Since we want to track not only the material flow but also the information and cost flow in order to obtain metrics and performance indicators, we propose to store all the simulation data in a database which leads us to save the parameters and input data for the initialization and the simulation execution.

To better explain the application of VCOR to our chain we started from the level 1 of the model, i.e. the Top Level, following a top-down approach. Once defined the Macro-Processes involved in each element of the chain, we choose its configuration and later we depict the level 3 elements implicated in the process. To describe the model we will use the following example (figure 3).

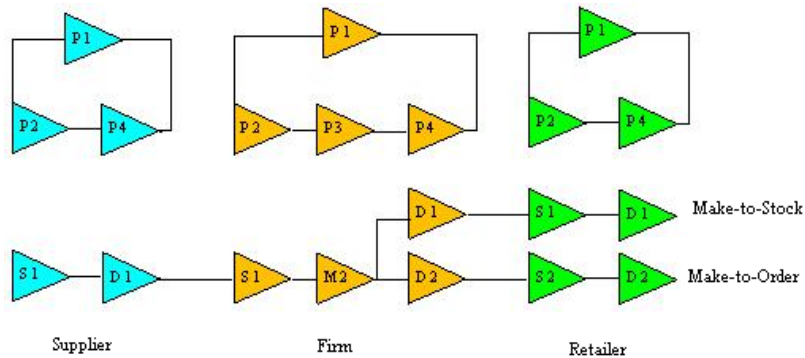

Figure 3 - Supply Chain Level 2 Configuration

We started from the central element of the chain: only the Firm has a production process that transforms raw materials in finished goods (3 products). For this reason the Firm is the only participant that has a *Make process* in its configuration. The other members, like Retailer and Suppliers, have only package processes that can be simply included in the *Deliver process*. For some products the production follows the rule of the *Make-to-Stock*, whilst for other products a *Make-to-Order* production is assigned. These considerations are summarized in the figure 3: the processes P1, P2, P3 and P4 represent the development and the establishment of activities over specified time periods that represent a projected appropriation of supply chain resources (materials, production, and delivery) to meet supply chain requirements.

One of the most important aims of the VCOR model is to describe and manage information, costs and material flows. In order to achieve this goal the simulation tool is based on the orders management. When any member needs a certain quantity of products it becomes a *client* and sends an order signal to its predecessor (i.e. the *provider*) in the chain. An order is characterized by its identifier, quantity of product, product type, timestamps and status. When a client makes an order, it updates the Demand database of its provider. Every time there is a change in the status of the order, the information is updated in the database. This mechanism allows us to exactly know the real position of commodities, monitoring both the material and the information flow.

To build our simulation model we used the discrete event oriented ARENA software; the necessary SCOR/VCOR Level 3 elements have been realized in ARENA blocks and then gathered and organized in sub-models in order to set up the different processes. In the following paragraphs we describe, as example, the implementation of different actors of the chain in the model:

*a. Retailer:* the retailer is composed by a Source and a Deliver process with two different configurations: *Make-to-Stock* and *Make-to-Order*. An Arena block is implemented with a rule which involves that each request made by the consumer becomes an order of the Retailer to the Firm. In the figure 4 some elements of the Source process are shown. The element called *Schedule Product Deliveries S2.1* has the role to check periodically the three inventory levels stored into the database. If the effective level is under a "s" value (chosen by the managers), this module sends an order to the Firm, writing information as the "*order id*", "*client id*", "*quantity*", "*product type*", etc. in the Firm's Demand database.

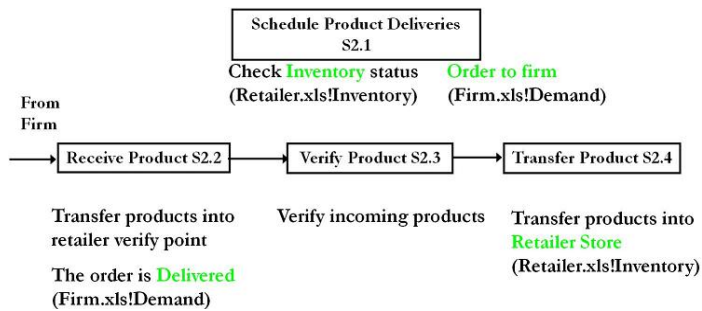

Figure 4 - Retailer VCOR Source process

With the application of VCOR model, the Retailer configuration defines a new process: the *Support process* which has the objective to solve the Customers problems when a delivered lot is defective because of the transport or a bad installation. The figure 5 illustrates this process.

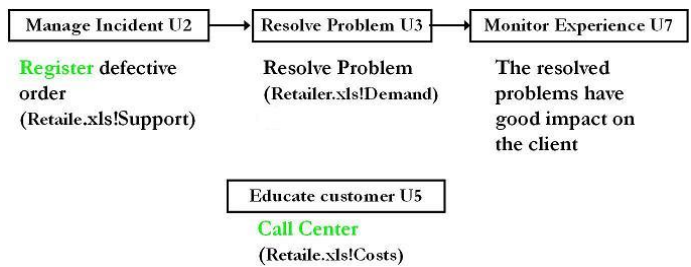

Figure 5 - Retailer VCOR Support process

The *Manage Incident U2* has the role to register in the Support database a defective order, the related quantity of products returned, the customer identifier and the timestamps. The *Resolve Problem U3* resolves the problem writing a new order in the Retailer Demand base with a quantity equal to the number of damaged products. Once the order is delivered the *Monitor Experience U7* registers the operation and modifies a variable in the Customer Behaviour model that increments the vote. The *Educate Customer U5* simulates a Call

Center, it has the important task to decrement the percentage of defective products decrementing the value of a variable in a Customer ARENA module.

*b. The firm:* The Firm has the role to transform raw materials coming from Suppliers into products for the Retailer. Its configuration includes a Source, a Deliver and a Make process. The following figure describes the ARENA representation of the "A5- Receive Order" process

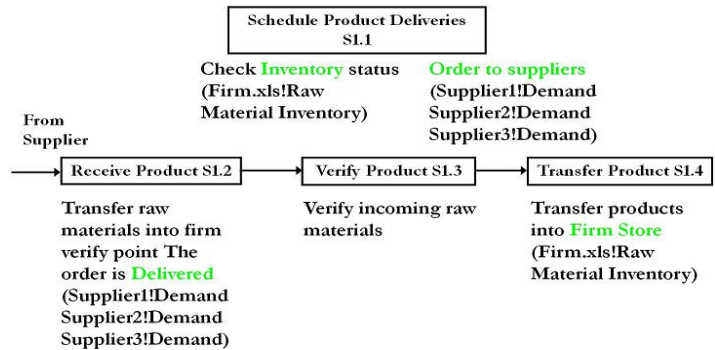

Figure 6 – Firm's Receive Order process

The Firm has also four processes derived by PLM and CRM of the Value Chain Management. The Market, Research and Develop processes belong to PLM, while the Sell and the Support to the CRM. With these new processes a company tries to analyze the market and takes consequently strategic decisions, like the restyling of a product or the acquisition of new technologies. Since it is very difficult to reproduce a complex market analysis or a restyling of a product, the ARENA blocks simulate these events in terms of required time and associated costs.

The *Analyze Market M1* (figure 7) periodically checks the Customer satisfaction level in the Retailer Demand database and, if the vote goes down under a specified threshold, it activates the *Architect Solution M4* module. This block finds the product type that has the least number of sales and decides to adapt it. In order to change a product, the Firm needs to modify its production line with the introduction of new technology. Once acquired the new equipment with *R3* and *R7* modules, *Introduce Technology R8* changes the production process in the VCOR *Build Product B3* that corresponds to the *M2.3* element.

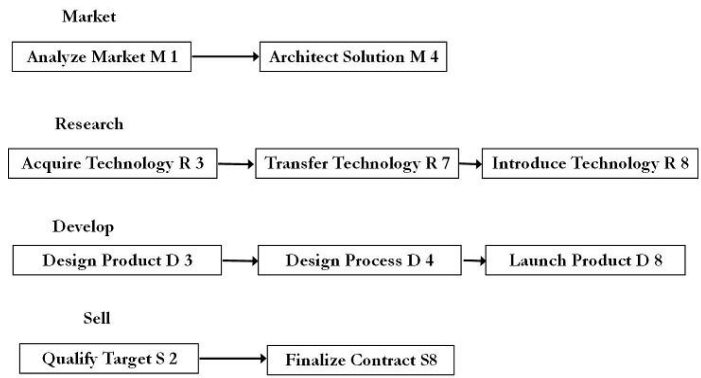

Figure 7 - Firm Market, Research, Sell and Develop processes

The *Develop process* materially changes the Bill of Material of the product and it is responsible to launch the product. In fact, the *D8* sets the variable "Innovation Factor" in the Customer Behaviour model and, as consequence, the customer's vote increases considerably. The *Sell process* tries to identify the clients in the market in order to develop relationships and proposals. The *Qualify Target S2* classifies the Clients on the basis of their priority and determines which of them can be supplied by the Firm. It calculates the total number of products required by the Clients and finalizes the contracts (*S8*) that must not exceed the fixed percentage of the dedicated production capacity.

### 4.2. The Customer Behavior model

In the simulation model, the Customers are able to express an opinion, in the form of a vote, which represents the satisfaction level for each single order delivered. This is necessary because it is at the base of the Value Chain philosophy in which the main goal is to give to the market what it needs, obtaining a feedback for the improvements introduced on the product and/or for the introduction of new products. For this reason the VCOR simulation needs a consumer behaviour model. Some models used questionnaires (Chen and Hughes, 2004), some others applied complex neural networks (Wen-Bao, 2007). In our simulation approach we have adopted the Adaptive Learning Model (Hopkins 2006), which consists in a mathematical model which is able to simulate the satisfaction level of the consumer and to pass the data to the corresponding simulation blocks.

According to the Adaptive Learning Model two customers generally express their level of satisfaction through a vote for each type of product. This vote decrements its value in exponential way which is coherent if we consider that a product has a life cycle and that at the end it becomes obsolete. The consumer satisfaction can increase if the support service, introduced in the VCOR example, is able to resolve the clients' problems. The factor that can drastically increment the customer satisfaction level is the introduction of a new product as result of a market analysis. This is possible in the mathematical model thanks to an "*Innovation Factor*" that is enabled when an old product is modified to meet

the demand of the market. Our model differs from the Hopkins's one for the following two main aspects:
- the introduction of a *forgetting factor α* in place of the coefficient *δ*, in order to model the behaviour of the customer to "forget" the improvement on a product with the time;
- the input function *u_i(k)*, which in the original model expresses the input of the customer, has been modified taking into account the following variables:
  – The product cost variation *Δp*: expressed in percentage, according to the market rules.
  – The delay *d*: expressed in percentage with respect to the delivery lead time of the product.
  – The Quality parameter *q*: expressed as a ration of conform products over the total product purchased (KPI).

The adopted consumer behaviour model can be now expressed with Hopkins notation, taking into account our modifications:

$$x_i(k+1) = (1-\alpha) \cdot x_i(k) + \alpha \cdot u_i(k) \quad (1)$$

Where:
  $x_i(k)$ : the vote at instant *k* of the *i* customer.
  $\alpha$: forgetting factor ($0<\alpha<1$)
  $u_i(k)$ the user input function
  In our model $u_i(k)$ has the following expression

$$u_i(k) = f(k) \cdot I_n + \xi \cdot s + \beta \cdot \Delta p + \delta \cdot d + \phi \cdot q + \eta \cdot x_{j \neq i}(k) \quad (2)$$

Where:
*s* indicates if the last request of support was correctly satisfied or not.
*ξ* takes into account how important is the quality of service.
$I_n$ is a Boolean parameter which indicates if the product delivered is a new one (1) or not (0). It changes its status to one when a new product is introduced in the market and to zero after a first order for this product is delivered. The purpose of the parameter is to increment the customer's vote.
$y_i(k)$ is the output of the system (i.e. the vote): $y_i(k) = x_i(k)$
*f(k)* is a specific function depending on the vote given at *k* time. It is based on the specific industrial context and depends essentially on the customer satisfaction detection system adopted by the firm in its Quality Management System. In our application case a measurement system will be adopted with respect to the firm's customer satisfaction requirements.

## 5. The case study

Our value chain simulation model has been tested on a supply chain related to an agro-food main firm of the southern Italy producing fruit and tomato juices. This firm is involved in a supply chain and uses an Enterprise Resource

Planning system (ERP) to integrate the information systems of its different sites. In this case study, we propose different scenarios related to three types of products (figure 8) which represent the 30% of the entire production of the firm and have a market share up to 15%.

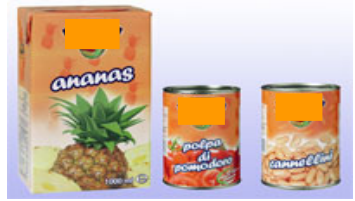

Figure 8 -The products chosen for the simulation test

Here the SCOR we be also used to model and analyze the different processes of the chain.

The firm receives fruits, vegetables and chemicals for product preservation from three different suppliers and provides their treatment, packaging and distribution to the retailers. The firm has a set of 10 big retailers that distribute the products to the final customers (supermarkets and small stores). In this study, we have selected only one Retailer with the highest number of customers (7 big supermarkets) and, among the customers, the two customers with the highest value of products sold (2 supermarkets that have a sale rate up to 20%). The supplier has one warehouse, while the firm and the retailer have two warehouses. The described supply chain can be structured as shown in figure 9, where the material flows can be seen too.

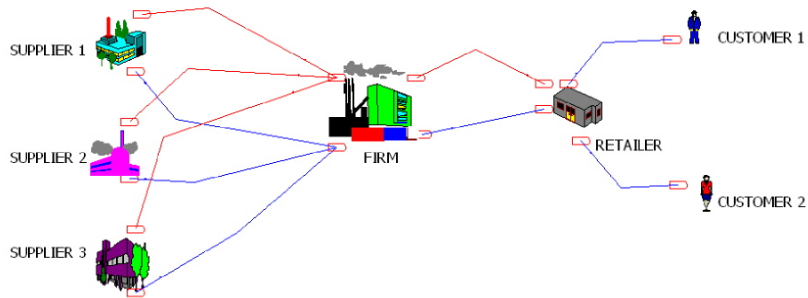

Figure 9 - Supply Chain structure

The *Supplier 1* is specialized in the production of only one component (named *Raw 1*), the *Supplier 3* for the *Raw 3* component, while the *Supplier 2* can supply the other two types of components (*Raw 1* and *Raw 2*) The entire configuration of raw materials distribution necessary for the production of finished goods is given in the following table:

**Table 2**   Production of raw materials by suppliers

| *Production of raw materials* | Supplier 1 | Supplier 2 | Supplier 3 |
| --- | --- | --- | --- |

|       |   |   |   |   |
|-------|---|---|---|---|
| Raw 1 |   | X |   |   |
| Raw 2 |   |   | X |   |
| Raw 3 |   |   | X | X |

Based on a first analysis we obtained the data related to the demand of the three described products. To do this we have monitored the demand of the two customers during a period of one year, divided in observation periods of one month; the data retrieved are shown in tables 3 for the customer 1 and 2, in which $M_i$ are the observation months. The reported values are in number of boxes for each product. We can see that for the product 1 the demand can be assumed as constant, while it is variable in a seasonal behavior for products 2 and 3.

**Table 3** Product demand for the Customer # 1 and Customer # 2

| *Dem. 1* | $M_1$ | $M_2$ | $M_3$ | $M_4$ | $M_5$ | $M_6$ | $M_7$ | $M_8$ | $M_9$ | $M_{10}$ | $M_{11}$ | $M_{12}$ |
|---|---|---|---|---|---|---|---|---|---|---|---|---|
| Prod. 1 | 250 | 260 | 245 | 247 | 255 | 257 | 250 | 251 | 253 | 255 | 250 | 241 |
| Prod. 2 | 550 | 659 | 580 | 650 | 770 | 850 | 890 | 790 | 700 | 650 | 590 | 500 |
| Prod. 3 | - | - | - | - | - | - | - | - | - | - | - | - |
| *Dem. 2* | $M_1$ | $M_2$ | $M_3$ | $M_4$ | $M_5$ | $M_6$ | $M_7$ | $M_8$ | $M_9$ | $M_{10}$ | $M_{11}$ | $M_{12}$ |
| Prod. 1 | 300 | 310 | 312 | 295 | 311 | 320 | 301 | 305 | 313 | 300 | 295 | 297 |
| Prod. 2 | - | - | - | - | - | - | - | - | - | - | - | - |
| Prod. 3 | 70 | 165 | 140 | 145 | 250 | 355 | 397 | 410 | 380 | 371 | 280 | 210 |

.

### 5.1. SCOR/VCOR model Simulation

The validation of the simulation model has been conducted with an experimental campaign, before with SCOR model and then with the extension of the VCOR ARENA modules. This section presents the most important results of the experiments conducted and some comparisons, with the real data of the firm. The unit time of the run is expressed in hours and the run length is two days. The most important parameters of the participant are shown in the table 4, where the numbers in square parenthesis refer respectively to product 1, 2 and 3.

**Table 4** Parameters Value used in the simulation

| *Participant* | *Parameter* | *Value* |
|---|---|---|
| Retailer | Products Inventory Level [1, 2] | [0, 0] |
|  | Rescheduling Frequency Deliver | 2 hours |
|  | Rescheduling Frequency Source | 2.5 hours |
| Firm | Finished Goods Inventory (FGI) | [500, 500, 300] |
|  | Raw Inventory Level (Kg) | [200, 200, 200] |
|  | Rescheduling Frequency Deliver | 2.5 hours |
|  | Rescheduling Frequency Source | 3 hours |
|  | Rescheduling Frequency Make | 3 hours |
|  | Max Daily Production Capacity | 185 |
| | FGI Inventory Level | 500 |
| *Suppliers* | Rescheduling Frequency Deliver | 4 hours |
| | Rescheduling Frequency Source | 4 hours |

Figure 10 illustrates the Retailer Delivery Time obtained by the run of the simulations. It is possible to note the peak at the 11th order (18 hours) is caused by the lead time for the replenishment of inventory; the mean time is 5.32 hours.

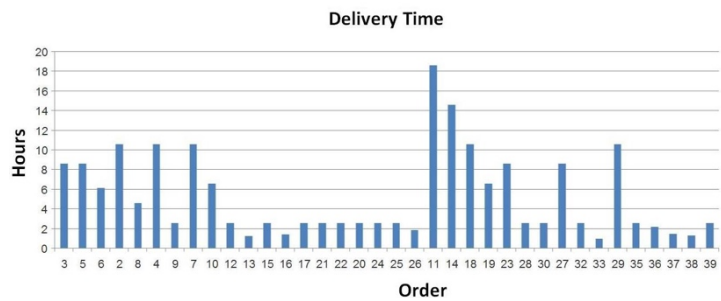

Figure 10 - Retailer Delivery Time

The following figure represents how the simulator has calculated the delivery time for the Firm. There are few orders delivered because at the end of the simulation some orders were in Finished Goods Inventory (FGI), "Open" or "In Transit" status. The mean time is rather low, due to the FGI initial stocks.

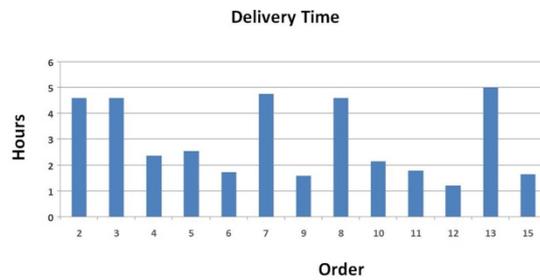

Figure 11 - Firm Delivery Time (mean time: 2.97 hours)

We have compared the data obtained with some historical data of the firm, obtaining a satisfactory congruence with the results of the simulation, which difference with real data has never been less than 12% in the worst case. As a further experimental validation of the model, we have noticed that during the entire simulation, the supplier number 2 has never sent an order to its supplier because of the Inventory Level calculated by the simulator which was enough to satisfy the Firm's demand. In fact, a survey conducted on the warehouse holding costs of the supplier number 2 has shown that they are up to the 87% of the total supply chain costs.

### 5.2. VCOR specifications and Customer Satisfaction Scenario: Simulation Results

The VCOR specification has been added through the Market and the Research processes. In this case we have made a previous survey related to the customer satisfaction level with the previous model for the product 1 (the one with the highest numbers of sells). The data have been acquired through the questionnaires of customer satisfaction distributed to the customers during the period in which the orders have been set. Figure 12 shows the results obtained, in which we can see a progressive decreasing of the vote related to the product with the increasing of the number of orders.

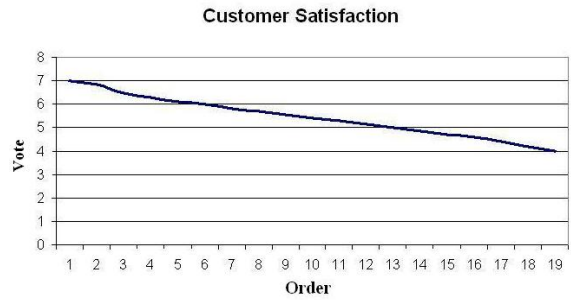

Figure 12 - Customer Satisfaction trend with SCOR

The extension to the VCOR has been made through a particularization of the Customer behavior model presented in section 4.2; in our model the function (2) related to the vote at instant *k* has a particular expression for the *f(k)* relation, in order to accomplish the specification of the votes found on the Customer Satisfaction modules used by the firm in its Quality Management System. In this application case the relation $u_i(k)$ has the following expression:

$$u_i(k) = [9 - 6/5 \cdot (1-\alpha) \cdot x_i(k)] \cdot I_n/\alpha + \xi \cdot s + \beta \cdot \Delta p + \delta \cdot d + \phi \cdot q + \eta \cdot x_{j \neq i}(k)$$
(3)

in which the notations are the same of equation (2) with the following additional considerations:

- The function *f(k)* has its values included into the range [0-10], according to the customer satisfaction system of the firm
- The $\Delta p$ variation has to be expressed as a percentage of the previous price and the weight $\beta$ has to be positive
- The delay *d* has to be expressed in percentage with respect to the mean lead time of the product.
- The Quality level *q* has to be expressed in percentage with respect to the mean quality level.

The simulation model has been run for the subsequent of the 19 orders. The figure 13 shows the customer satisfaction level in the simulation with VCOR model from the order number 20 to the order number 39, including the Support process in the Retailer and the whole PLM part in the Firm. The simulation has generated a high elevation of the trend after the 20$^{th}$ order due to the introduction in the market of the renewed product (a new taste and nutritional features have been added due to surveys conducted to the customers with and additional survey data base introduced by the CRM module).

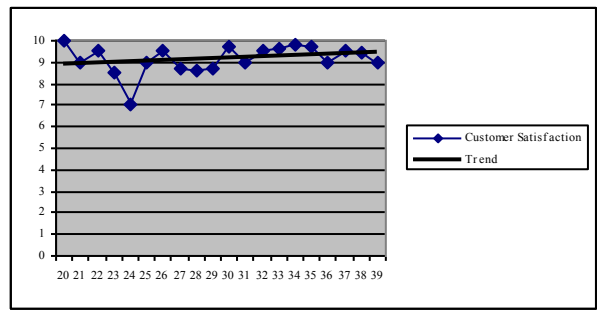

Figure 13 – Customer Satisfaction trends after VCOR simulation

As a further validation of the proposed simulation model, a post-sales survey has been conducted on the product for which the simulation has been run. In Figure 14 we can see the real data, coming from the above mentioned Customer Satisfaction questionnaires. We can see that the values and the related trend reflect the simulation analysis with a maximum error of not more than 20% in the worst case.

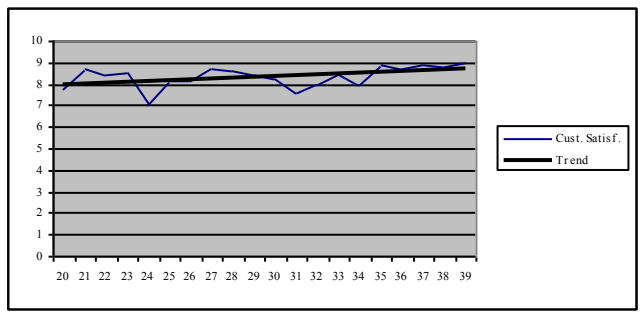

Figure 14 – The results of the survey on customer satisfaction

### 5.3. Delivery Times Scenario: Simulation Results

Another interesting result from the Value Chain simulation is related to the delivery times of the Retailer. In the histogram of figure 15 we report the simulation result of the entire period (from the first to the 37$^{th}$ order). With respect to the SCOR case we have noticed an increasing of the mean delivery time (7.84 hours); this increasing has been investigated inside the firm, finding the main two factors:

- With the introduction of the Sell process, the Firm decreases its finished goods inventory level to satisfy the new Customers demand, slowing down the Retailer's orders.

- The Support process of the Retailer increases the mean to 2.3 hours; it is due to the support time.

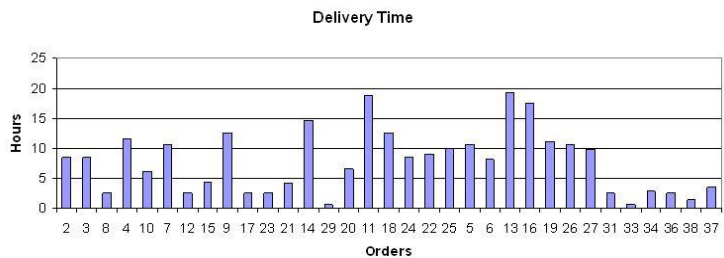

Figure 15 - Retailer Delivery Time

The Firm's behavior is more complex and for this reason two simulations have been necessary. The first one analyzes the Firm without the value chain part, in which the mean time is 5.36 hours (figure 16), and the second one investigates the whole performance.

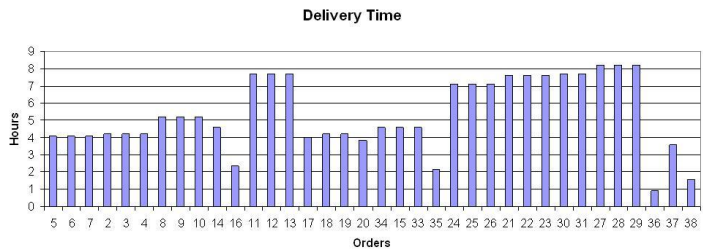

Figure 16 - Firm Delivery time

With the introduction of the Sell process in the Firm, we can notice that in the previous case the firm has satisfied about fifteen orders and now, with the application of CRM part of VCOR, more than thirty. The high increment is mainly addressed to the acquisition of new customers that guarantees fixed financial entrances, to which Retailer's demand must be added. Moreover, the production increases from 296 to almost 500 finished products, stressing the productive cycle to higher level. To sum up, there are some benefits deriving from the adoption of the VCOR model. To better understand the improvements obtained by this choice, some indicators have been calculated to test the performance of the chain:

- *Stock Rotation Indicator:* it shows the sale rapidity of the commodities. It is calculated as:

$$SRI = \frac{Sales\ Profit}{Mean\ value\ of\ the\ stock}$$

in which a big value indicates the capability of the company to use our products.

- *Stock Mean Time:* it indicates the mean time of undelivered materials or goods. It is calculated as:

$$SMI = \frac{Period}{SRI}$$

where *Period* is the simulation timeline, expressed in hours.

- *Sales Profitability Indicator:* it gives an idea on how the costs influence the effective profit. It is calculated as:

$$SPI = \frac{Sales\ Profit - Costs}{Sales\ Profit}$$

In table 6 the enhancements can be seen; we can notice a huge improvement in all of the Firm's indicators, especially for the *SMI* for the finished goods and the raw materials inventory.

**Table 6** Retailer's Performance Indicator

| Indicator | VCOR | SCOR |
|---|---|---|
| *SRI* for FGI | 22.4 | 12.34 |
| *SMI* for FGI | 2.13 hours | 3.88 hours |
| *SRI* for Raw Materials | 1.2 | 0.69 |
| *SMI* for Raw Materials | 39.8 hours | 68.8 hours |
| *SPI* | 13% | 12% |

This last indicator underlines that with the VCOR a lot of commodity is used (sold or used for the production) in the 45% of the time required with the SCOR simulation. We can also see that the *SPI* remains almost unchanged. However, if we consider that in the second simulation there are also the Support costs, the fact that the *SPI* is constant means that the Support costs do not affect much the total costs of the Retailer.

## 6. Conclusions

This paper has discussed the study of the two most known models used to implement the concepts of Supply Chain Management and Value Chain Management, SCOR and VCOR, through a simulation approach. Starting from an analysis of their standard architectures, using a bottom-up approach, and their performance indicators and metrics, a simulation framework has been developed under ARENA. Once implemented the SCOR model the tool was extended to VCOR.

The simulation models have been applied into an agro-food firm in which the main experimental results given by the simulation runs have been compared with the real data of the firm. This has been done to assess the model proposed and to verify the difference between the simulated data and the real ones. As confirmed by the results of these simulations, the adoption of VCOR model needs of bigger financial and organizational efforts, but these efforts can be fully repaid by the benefits in terms of quality of service, market extension, competitiveness, flexibility, "quick response", innovation and other features essential for the firms to survive in the global market.

Finally, the results of some simulations have been presented, including a comparison of the two models in terms of the main *Key Performance Indicators* that are crucial for the supply chain analysed. A future work could complete the implementation of the templates to all the processes of VCOR model, extending the flexibility and the reusability of the realized tool.